\documentclass[
twocolumn,
superscriptaddress,
amsmath,amssymb,
aps,
]{revtex4-2}
\usepackage{graphicx}
\usepackage{hyperref}
\usepackage{dcolumn}
\usepackage{bm}
\usepackage{epstopdf}
\usepackage{romannum}
\usepackage{csquotes}
\usepackage[dvipsnames]{xcolor}
\usepackage{hyperref}
\usepackage{float}
\usepackage{graphicx} 

\usepackage{epsfig}
\usepackage{epstopdf} 
\usepackage{afterpage} 

\begin{document}
	
\preprint{}
	
\title{Strong correlation between $H$-linear magnetoresistance and strange metal in FeSe superconductor}
	
\author{Xinyue Wang}
\affiliation{School of Physics, Southeast University, Nanjing 211189, China}
\author{Yue Sun}
\email{Corresponding author: sunyue@seu.edu.cn}
\affiliation{School of Physics, Southeast University, Nanjing 211189, China}
\author{Wei Wei}
\affiliation{School of Physics, Southeast University, Nanjing 211189, China}
\author{Qiang Hou}
\affiliation{School of Physics, Southeast University, Nanjing 211189, China}
\author{Nan Zhou}
\affiliation{Key Laboratory of Materials Physics, Institute of Solid State Physics, HFIPS, Chinese Academy of Sciences, Hefei 230031, China}
\author{Yufeng Zhang}
\affiliation{The School of Physics and Electronic Engineering, Jiangsu University, Zhenjiang 212013, China}
\author{Zhixiang Shi}
\email{Corresponding author: zxshi@seu.edu.cn}
\affiliation{School of Physics, Southeast University, Nanjing 211189, China}

\begin{abstract}
	In strange metals, a strong and anomalous scattering effect exists and increases linearly with temperature. In FeSe, we observed that the temperature dependence of resistivity exhibits non-Fermi liquid behavior in two regions below and above a critical pressure, $p_\text{c}$\,$\sim$\,2\,GPa. As pressure increases, a transition from quadratic to nonsaturating magnetoresistance is observed, with a distinct crossover between these two behaviors indicated by $B^{*}$ in the derivative analysis. After subtracting the quadratic term from the magnetoresistance, the residual magnetoresistance clearly exhibits an $H$-linear behavior. Additionally, two segments of $H$-inear magnetoresistance appear with increasing pressure, each arising from distinct origins. Notably, the two $H$-linear magnetoresistances exist and develop within the strange metal states. These results suggest that $H$-linear magnetoresistance is in strong correlation with the strange metal state, which may affect superconductivity in FeSe under pressure. Our study provides valuable insights into the strange metal state and clues for underlying unconventional superconductivity.
		
\textbf{}

\end{abstract}
	
	
\maketitle

 Strange metals exhibit highly unconventional electrical characteristics, where, in contrast to Landau’s Fermi liquid theory, non-Fermi liquid\,(NFL) behavior is applicable \cite{WOS:000277834300349}.
 Despite the lack of a unified theoretical description of strange metals, different systems exhibit some similar behavior. For example, NFL phenomena associated with linear resistivity have been widely observed in the phase diagrams of unconventional superconductors \cite{WOS:000463851300002,jie,19,Ba2024,crossFSS,Ba122nihe,PhysRevResearch.3.023069,WOS:000316290200002,PhysRevLett.101.136408,WOS:000184859300027}. 
 Jie Yuan et al. reported a quantitative scaling law among $T_\text{c}$ and the linear-in-$T$ scattering coefficient in electron-doped copper oxide La$_{2-x}$Ce$_x$CuO$_4$ \cite{jie}. This discovery suggests an interaction between high-temperature superconductivity and the strange metal state. The transport properties of electrons in the strange metal state may be related to the pairing in the superconducting state. Subsequently, Shu Cai et al. reported a similar quantitative relationship in iron-based superconductors (IBSs) under high pressure \cite{19}. Additionally, recent studies on Bi2201 have further established a close relationship between the concentration of carriers in the strange metal state and superconductivity \cite{Ba2024}. 
 These findings help to connect the strange metal state with unconventional superconductivity, revealing a possible common mechanism behind them.

 In strange metal regions, quantum critical fluctuations are very common \cite{WOS:000377670000004,WOS:000400818400034}.
 Although quantum phase transitions occur at zero temperature, its fluctuations can affect a considerable non-zero temperature region. 
 Unique transport phenomenas and unusual scaling behaviors emerge in this region \cite{WOS:000316290200002,PhysRevLett.101.136408,WOS:000184859300027}. 
 Recently, quasilinear magnetoresistance (MR), which possibly originating from a strange metal component, has been detected in BaFe$_2$(As$_{1-x}$P$_x$)$_2$ \cite{Ba122nihe}, Bi2201 \cite{Tl2201} and FeSe$_{1-x}$S$_x$ \cite{PhysRevResearch.3.023069}. The $H-T$ scaling of the MR observed in the NFL region would be a method to predict the potential quantum critical fluctuation.
 Inspired by the observation of linear MR above, some have proposed that this behavior represents another aspect of Planckian dissipation in strange metals \cite{WOS:000822228900036}. Specifically, the scattering rate would exhibit an anomalous not only on $T$-linear dependence but also on $H$-linear dependence.

 Meanwhile, in unconventional superconductors, antiferromagnetism\,(AFM), spin-density wave\,(SDW), and nematicity are commonly observed in the region of strange metals \cite{WOS:000262862800034,WOS:000914347400007,WOS:000543777300003}. These phases significantly affect both the superconducting and normal state properties. According to Landau's theory of phase transition, each phase can be described by an order parameter, which can be adjusted through doping, gating, and applying pressure.
 However, up to now, there is still no consensus on the relationships between these ordered phases and their connection to superconductivity. For example, some researches suggest that the nematic phase is a precursor to the AFM phase, breaking discrete lattice rotational symmetry before breaking continuous rotational symmetry \cite{PhysRevB.78.020501,PhysRevLett.107.217002,WOS:000306505100006}. On the other hand, from the perspective of orbital degrees of freedom, the nematic phase is interpreted as a type of orbital-ordered state \cite{PhysRevB.80.224506,PhysRevB.79.054504,OrbitalMR}, distinct from the AFM phase. Experiments indicate that AFM spin fluctuations are strongly enhanced toward $T_\text{c}$, suggesting a positive link between the AFM fluctuations and superconductivity \cite{PhysRevLett.102.177005}. Besides, a competitive relationship between magnetism and superconductivity are also proposed \cite{crossFSS}.
 In IBSs, AFM is observed always accompanied by a structural transition from tetragonal to orthorhombic, or more precisely, a nematic phase transition \cite{WOS:000890204900001}. This characteristic complicates the study of each phase. Therefore, the relationship between these orders, and strange metal behavior is a crucial issue in superconductors and key to understanding their unique physical properties.  

 \begin{figure}[t]\centering
	\includegraphics[width=1\linewidth]{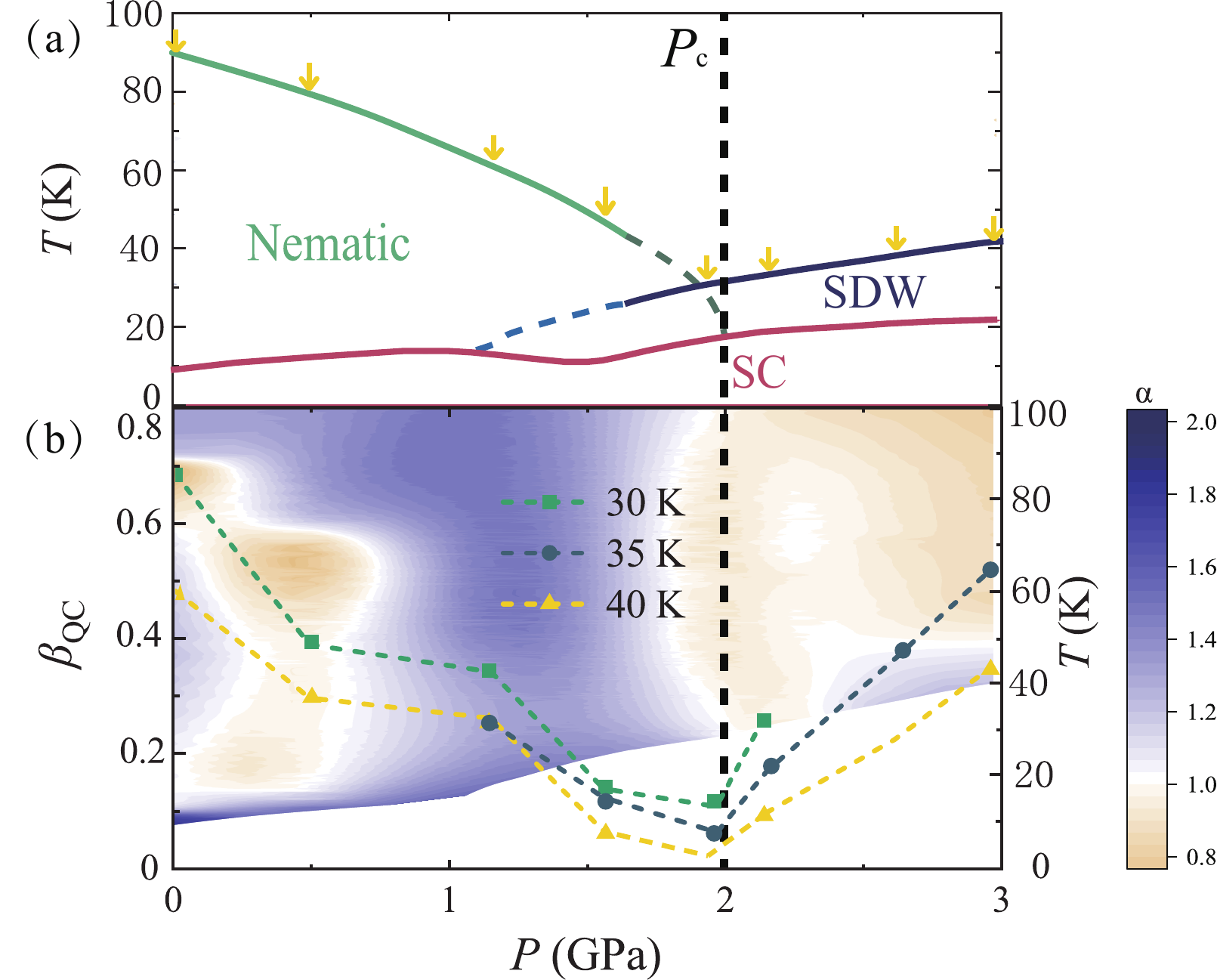}
	\caption{(a) Pressure-temperature ($P-T$) phase diagram of bulk FeSe. The nematic ($T_\text{s}$, green), magnetic ($T_\text{m}$, blue), and superconducting transition temperatures ($T_\text{c}$, red) as a function of pressure are from previous work \cite{SJP}. The dashed line indicates the possible existence of nematic and magnetic phase and guide to the eyes. The yellow arrows indicate the pressure measured in the current work, and the resistivity behavior is consistent with the ref. \cite{SJP}.
	(b) Contour plot of the temperature dependent exponent $\alpha$ obtained from $dln(\rho-\rho_0)/dlnT$ and $\beta_\text{QC}$ (left axis) be extracted from $d\rho/d(\mu_0H$) at 30\,K, 35\,K, and 40\,K in FeSe. The data were obtained from measurements conducted in the PCC from this work.}    
	\label{FIG. 1}
\end{figure}

\begin{figure}[t]\centering
	\includegraphics[width=1\linewidth]{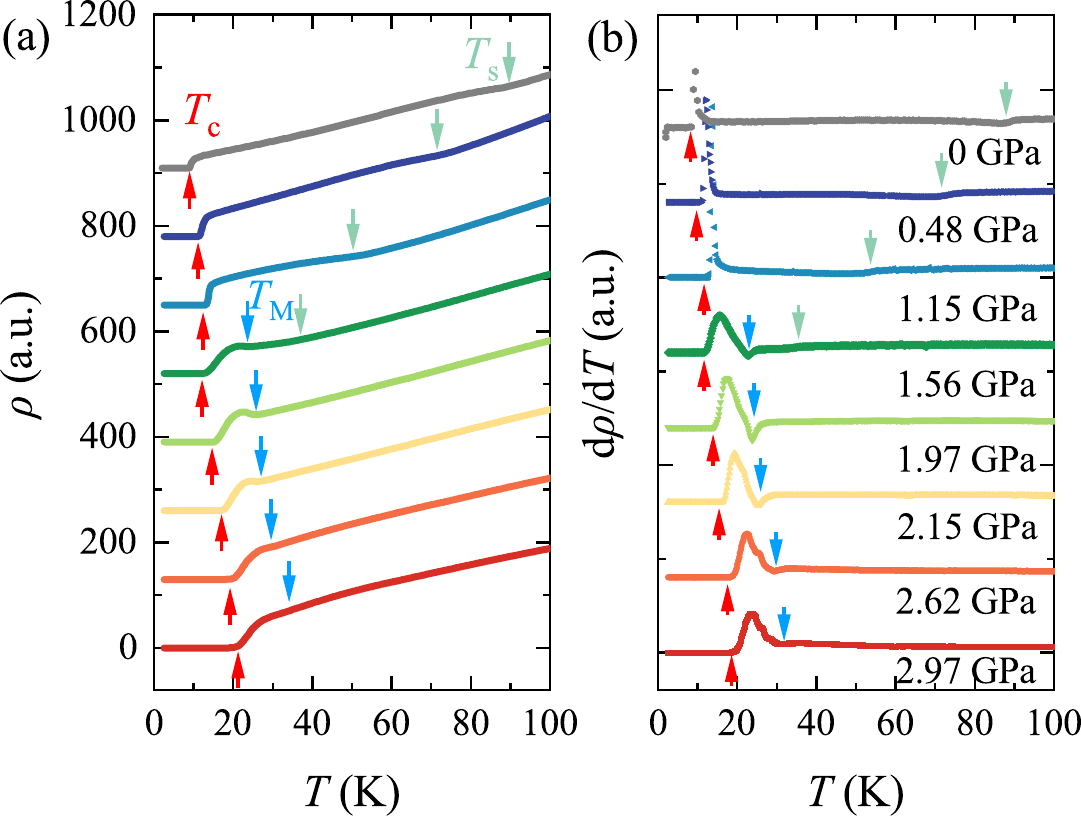}
	\caption{(a) Temperature dependence of resistivity, (b) Temperature dependence of $d\rho/dT$ below 100\,K for FeSe under various pressures. The data are vertically shifted for clarity. The arrows indicate the nematic transition temperature $T_\text{s}$ (green), magnetic transition temperature $T_\text{m}$ (blue), and superconducting transition temperatures $T_\text{c}$ (red).}    
	\label{FIG. 5}
\end{figure}

 FeSe is a unique case for studying the strange metal behavior as well as these ordered states \cite{HQ,fernandesWhatDrivesNematic2014}. It exhibits a strange metal state, and a nematic transition without long-range AFM under ambient pressure \cite{PhysRevLett.104.087003}. Under low pressure, the SDW order appears and the nematicity intertwines with SDW \cite{WOS:000418153100001,WOS:000368766100016}. 
 However, as shown in Fig. 1(a), at $p \textgreater$ 1.56\,GPa, the two orders decouple \cite{PhysRevLett.123.167002}, leading to a fixed magnetic ground state and the gradual emergence of long-range magnetic order.
 To study the relationship between strange metal, superconductivity, and other ordered states, we studied the MR of FeSe single crystal under pressures of 0, 0.48, 1.15, 1.56, 1.97, 2.15, 2.62, and 2.97\,GPa as marked by the arrows in Fig. 1(a). We divided the MR into two components, linear and quadratic terms. 
 Notably, the linear MR is just present in the regions of strange metal, and its magnitude, $\beta_\text{QC}$, spontaneously evolves with the change in the slope of resistivity, which suggests that the linear resistivity and linear MR are two sides of the same coin for the strange metal state.
\textbf{}
\begin{figure*}[htbp]\centering    
	\includegraphics[width=0.9\linewidth]{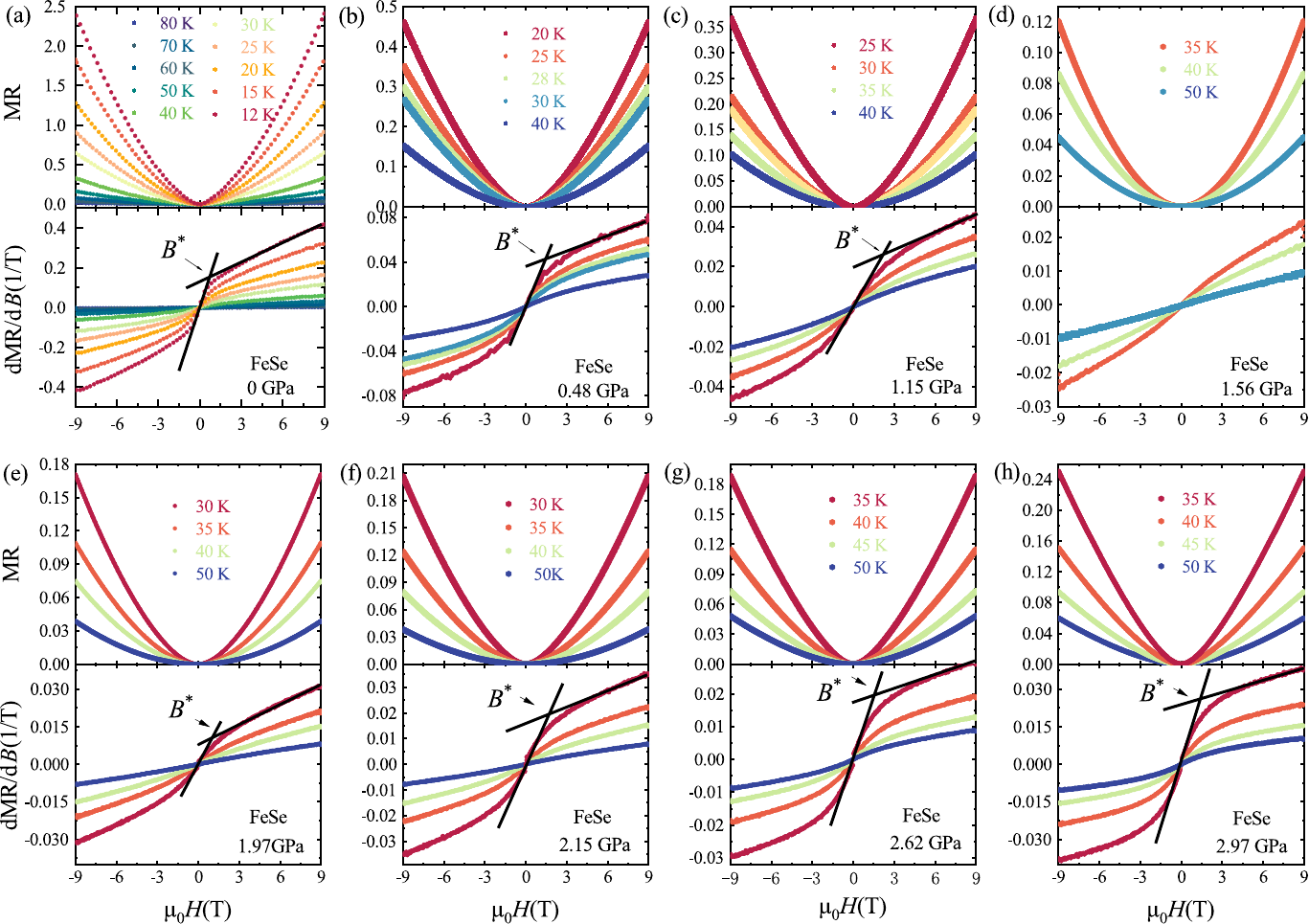}   
	\caption{Magnetic field dependence of {MR = ($\rho(H)-\rho(0))/\rho(0)$}, and field derivative MR [d(MR)/d$B$] for FeSe single crystals under (a) 0\,GPa, (b) 0.48\,GPa, (c) 1.15\,GPa, (d) 1.56\,GPa, (e) 1.97\,GPa, (f) 2.15\,GPa, (g) 2.62\,GPa, (h) 2.97\,GPa at different temperatures, respectively.}    
	\label{FIG. 2}   
\end{figure*}

 FeSe single crystals used for high pressure measurements were obtained by the chemical vapor transport growth \cite{FeSeCVT,PhysRevB.93.104502}.
 Quasi-hydrostatic pressure was applied via a clamp-type piston-cylinder pressure cell with Daphne oil 7373 as the pressure-transmitting medium \cite{gaoyafangfa}. 
 Transport measurements were performed by a physical property measurement system (PPMS 9 T, Quantum Design) with the pressure cell. Electrical resistivity measurements were carried out using a standard four-probe method in which the electrical contact was made by using silver epoxy. In order to calculate the applied pressure, a change in the SC transition temperature of Pb, which was contained together within the pressure cell, was resistively measured. The high-field transport measurement was carried out in a water-cooled magnet with steady fields up to 20 T at the High Magnetic Field Laboratory of the Chinese Academy of Sciences by using standard a.c. lock-in technique.

 The nematic transition temperature ($T_\text{s}$), magnetic transition temperature ($T_\text{m}$), and superconducting transition temperature ($T_\text{c}$) for FeSe under pressures are obtained from the resistivity measurements as shown in Fig.~\ref{FIG. 5}.
 To depict the evolutions of $T_\text{s}$, $T_\text{m}$ and $T_\text{c}$ with pressure, we construct the $P-T$ phase diagram in Fig.~\ref{FIG. 1}(a) with our data and those from ref. \cite{SJP}. 
 Under ambient pressure, FeSe shows a structural transition at $T_\text{s}\sim$90 K, changing from tetragonal to orthorhombic, with a nematic phase below $T_\text{s}$, determined by elastoresistance measurements \cite{92}, Raman spectra \cite{TsR} and high-pressure synchrotron X-ray diffraction \cite{TsXRD}.
 As pressure increases, the nematic phase is gradually suppressed, and long-range magnetic order forms at $\sim$1\,GPa. Since there is no obvious feature in the resistivity, this region is indicated by a blue dashed line. Around 1.5\,GPa, a noticeable kink appears in the resistivity, corresponding to the SDW transition temperature $T_\text{m}$, determined by $\mu$SR \cite{PhysRevB.97.224510} and NMR \cite{TmNMR}, shown as a blue solid line in the figure. When the pressure increases up to a critical value, $p_\text{c} \sim$ 2\,GPa, where the nematic phase is further suppressed, the transition temperature $T_\text{m}$ continues to increase. Above $p_\text{c}$, the structural and magnetic transitions occur sequentially \cite{PhysRevB.97.224510} and merge into a combined first-order-like transition proved by $\mu$SR measurement \cite{PhysRevB.97.224510}, indicating the presence of a magnetic tricritical point at $p_\text{c}$ \cite{PhysRevB.97.224510}. In addition, $T_\text{c}$ increases with pressure until $\sim$1\,GPa, followed by a dip when the nematic phase is fully suppressed. With further increase in pressure, $T_\text{c}$ rises with the enhancement of $T_\text{m}$. 

 The contour plot in Fig. 1(b) is obtained by fitting the temperature dependence of resistivity using the formula $\rho = \rho_0 + A \cdot T^\alpha$ in the range above $T_\text{c}$ and $T_\text{m}$ up to 100\,K. 
 The exponent $\alpha$ maintains a nearly constant value slightly below 1 in nematic region and around 1 near $T_\text{s}$. Outside the nematic phase, with increasing pressure, a dominant $T^{1.8}$ dependenceis visible at high temperatures, indicating an evolution from NFL to Fermi-liquid transport. Above $p_\text{c}$, the exponent $\alpha$ reduces to 1, which suggests the reappearance of NFL transport.
 
 To study the strange metals behavior, we measured the MR\,[=($\rho(H)$-$\rho(0)$)/$\rho(0)$] under pressures of 0, 0.48, 1.15, 1.56, 1.97, 2.15, 2.62, and 2.97\,GPa.
 As shown in Fig.~\ref{FIG. 2}, the value of MR gradually decreases when $p \textless p_\text{c}$ and then increases when $p \textgreater p_\text{c}$. 
 Besides, the slopes change can be seen more clearly in the d(MR)/d$B$ plot. As marked by the solid black lines, d(MR)/d$B$ linearly increases with magnetic field at small fields, which indicates a classic $B^{2}$ dependence. Above a characteristic field $B^{\ast}$, due to the contribution of the $H$-linear MR, the slope of d(MR)/d$B$ significantly decreases. Although our high-pressure measurements were limited to a magnetic field strength of 9 T, data from our high-field measurements under ambient pressure indicate that the $H$-linear MR can be sustained until 20 T (see more data in the SM \cite{SM}). The $B^{\ast}$, representing the crossover field, has been very weak at 1.56\,GPa as shown in Fig.~\ref{FIG. 2}(d). 
 Contrary to expectations, a distinct $H$-linear MR reappears at $p \geqslant $ 1.97\,GPa as shown in Figs. 3(e) to 3(h). 

\begin{figure}[t]\centering
	\includegraphics[width=1\linewidth]{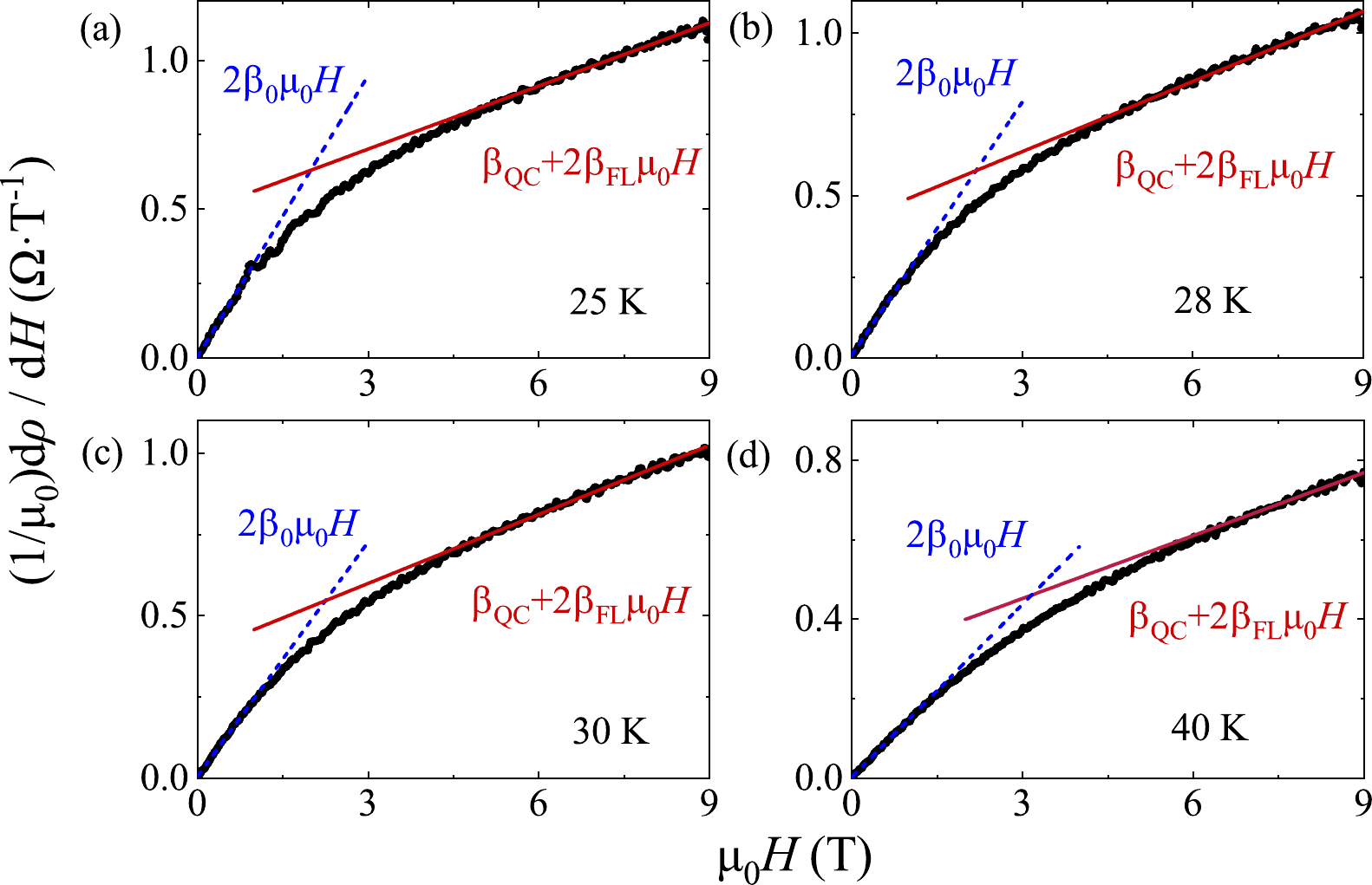}
	\caption{$d\rho/d(\mu_0H$) versus $\mu_0H$ at $p$\,=\,0.48\,GPa for (a) $T$\,=\,25\,K, (b) $T$\,=\,28\,K, (c) $T$\,=\,30\,K and (d) $T$\,=\,40\,K. The solid red lines and the blue dotted lines indicate the low-field $H^2$ dependence and the $H$-linear dependence, respectively.}
	\label{FIG. 3}
\end{figure}

 To separate and present the evolution of $H$-linear term with pressure more clearly, Fig.~\ref{FIG. 3} shows a typical results of $d\rho/d(\mu_0H$) at 0.48 GPa. The dashed lines represent the $H^2$-component. The solid lines are a linear fit from 7 to 9\,T and contains the contributions of the $H^2$- and the $H$-linear terms, which can be expressed as $\beta_\text{QC}+2\beta_\text{FL}·\mu_0H$, where $\beta_\text{QC}$ and $\beta_\text{FL}$ are the magnitudes of the $H$-linear and $H^{2}$- terms.
 We extracted $\beta_\text{QC}$ at 30\,K, 35\,K, and 40\,K from $d\rho/d(\mu_0H$) and ploted them in Fig.~\ref{FIG. 1}(b). 
 Surprisingly, $\beta_\text{QC}$ shows similar pressure dependent behavior as \( \alpha \). When $p \textless p_\text{c}$, $\beta_\text{QC}$ continuously decreases, while \( \alpha \) gradually increases from 1 to 1.4. After reaching a dip near $p_\text{c}$, $\beta_\text{QC}$ starts to rebound, indicating that the linear term contribution strengthens again. 
 \( \alpha \) also decreases to $\sim$1 after $p \textgreater p_\text{c}$. 
 Therefore, we infer that the $H$-linear MR is closely linked to the strange metal behavior. Similar results have been reported in BaFe$_2$(As$_{1-x}$P$_x$)$_2$ \cite{Ba122nihe} and FeSe$_{1-x}$S$_x$ \cite{crossFSS}. This result suggests that there is not only an anomalous $T$-linear dependence but also an anomalous $H$-linear dependence in the strange metal region.

\begin{figure*}\centering
	\includegraphics[width=0.8\linewidth]{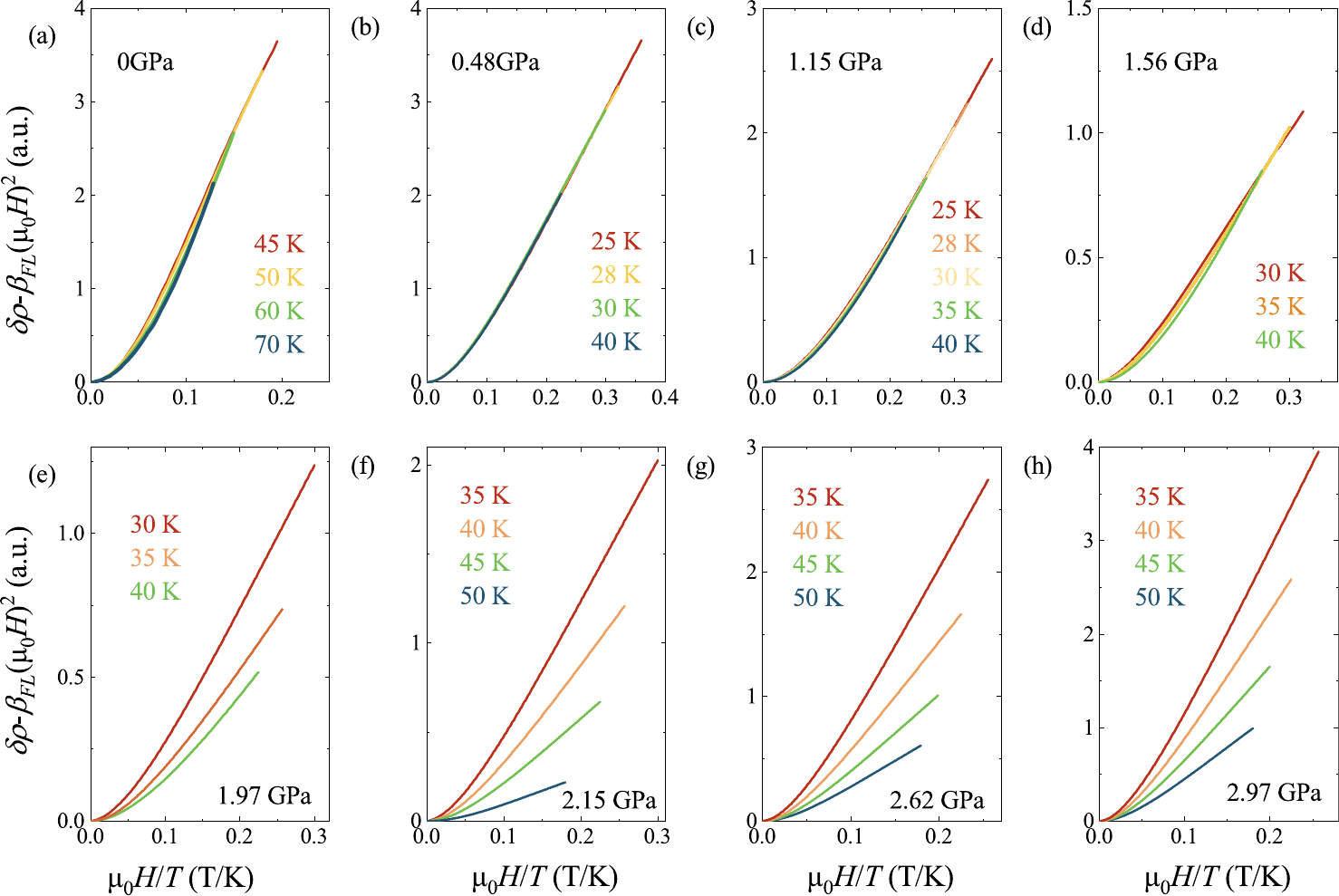}  
	\caption{$H/T$ scaling in the residual transverse MR in FeSe. Renormalized residual transverse MR in FeSe obtained by subtraction of $\rho$(H = 0) and the $\beta_\text{FL}(\mu_0H^2)$ orbital component at temperatures from 25 K to 70 K.}    
	\label{FIG. 4}
\end{figure*}

 Nevertheless, whether the two sections of $H$-linear MR above and below $p_\text{c}$ share the same mechanism remains to be discussed.
 Below $p_\text{c}$, the $H$-linear MR only exists within nematic region, which has been discussed previously from the Dirac-like band structure with ultra-high mobility as detected by APRES \cite{ARPES3,PhysRevLett.115.027006,PhysRevB.93.104502}. The Dirac band may come from the ferro-orbital ordering related to band shift, which disappears as the temperature rises above $T_\text{s}$ \cite{ARPES3}. 
 When the pressure increases up to 1.56\,GPa, the $H$-linear MR almost disappears.
 To further verify whether the $H$-linear MR is directly related to nematicity, we performed MR measurements on the H$^+$-intercalated FeSe (see more data in the SM \cite{SM}), where $T_\text{s}$ is completely suppressed, and $T_\text{c}$ is enhanced to over 40\,K \cite{HFeSe}. Clearly, only containing the quadratic term, MR of the H$^+$-FeSe confirms that the $H$-linear MR is directly related to the nematic order.
 In summary, the $H$-linear MR observed in Figs.~\ref{FIG. 2}(a) to 3(d) may arise from high-mobility carriers in the Dirac band and is directly related to the nematic order.

 To further check if the $H$-linear MR component below and above $p_\text{c}$ share the same origin, we applied the fitting method reported by Ian M. Hayes \cite{Ba122nihe}. 
 In this method, $\delta\rho-\beta_{FL}(\mu_0H)^2$ at different temperatures can be scaled into a straight line, where $\delta\rho=\rho(H,T)-\rho(0,T)$, and $\beta_{FL}$ is the quadratic coefficient. In addition to copper oxide superconductors, such scales also hold true in iron chalcogenides \cite{crossFSS} and heavy fermion superconductors \cite{CrAs}. Currently, the scaling is attributed to changes in electron scattering related to strange metal behavior near QCP. Fig.~\ref{FIG. 4} shows $H/T$ scaling of MR in FeSe under different pressure. Clearly, $\delta\rho-\beta_\text{FL}(\mu_0H)^2$ measured at various temperatures can be well scaled into a single curve when $p \textless p_\text{c}$, as shown in Figs.~\ref{FIG. 4}(a) to 5(d). 
 However, the scaling fails when $p \textgreater p_\text{c}$ as shown in Figs.~\ref{FIG. 4}(e) to 5(h).
 Clearly, the different scaling results before and after $p_\text{c}$ indicate that the origins of $H$-linear MR in the two regions are different. The failure of the scaling at $p \textgreater p_\text{c}$ may be due to the emergence of the long-range magnetic order \cite{SDW}. 

 There are several possible origins for the $H$-linear MR at $p \textgreater p_\text{c}$. One is the effect of order and disorder effect \cite{Tl2201,PhysRevB.82.085202}. However, high pressure is a relatively clean method introducing no additional disorder.
 Another possible reason is the complex carrier situation in a multiband system \cite{crossFSS,Tl2201}. Although there is a change in the dominant type of carriers between 2\,GPa and 3\,GPa \cite{CJG}, the band structure of FeSe does not show significant changes \cite{PhysRevB.90.144517}. Therefore, we can simply rule out the above two possibilities.

 Another possible origin of the $H$-linear MR at $p \textgreater p_\text{c}$ is related to the QCP.
 Similar scaling has been comfirmed near the QCP in iron chalcogenides \cite{crossFSS}, heavy fermions \cite{WOS:000238426600022}, the CrAs \cite{CrAs}, and most of the cuprate family \cite{Tl2201,cuprate2,cuprate3}. Although the mechanism of the scaling remains unclear, it is likely related to Planckian dissipation because this scaling works near the QCP \cite{WOS:000822228900036}.
 Apart from the common QCP, research on IBSs indicates the possible existence of the nematic quantum critical points (NQCP) in S- and Te- doped FeSe \cite{92,139}. Theoretical studies suggest that quantum critical fluctuations of the nematic phase may also lead to some anomalous behaviors in the normal state \cite{PhysRevLett.114.097001,171}. Since the nematic phase is driven by mechanisms such as orbital order, electron-phonon coupling, or electronic correlation effects, which are completely different from those of the SDW. Additionally, SDW decouples from the nematic phase at $p\,\sim$\,1.56 GPa and does not disappear until 6 GPa. Therefore, we primarily focus on the impact of the NQCP on the $H$-linear MR.
 In Fig.~\ref{FIG. 1}(b), near $p_\text{c}$, the nematic phase just vanishes \cite{PhysRevLett.121.077001}. According to previous studies on S-doped FeSe \cite{92}, the nematic transition temperature decreases with S-doping, while the nematic fluctuation is significantly enhanced. Ref. \cite{crossFSS} also found that with increasing disorders, the orbital MR is quenched, leading to the appearance of strict quantum critical scaling at or near the QCP.
 Hence, the linear MR may be caused by the nematic fluctuation, which could be enhanced at $p \textgreater p_\text{c}$.

 In addition, the density-wave formation is also discussed frequently as another possible reason of the linear MR as already been discussed in TaS$_2$, NbSe$_2$ and GdSi \cite{WOS:000470136000025,PhysRevResearch.4.033195,WOS:000470136000025,PhysRevLett.100.196402,PhysRevLett.102.166402,PhysRevB.94.235135}. 
 The presence of SDW under pressures above $p_\text{c}$ enhances the consequence of itinerant carriers turning at sharp corners of the Fermi surface. The formation of relevant electronic states opens a small curvature orbit at the Fermi surface. 
 Shubnikov-de Haas oscillation experiments show that there is indeed a very small sharp corner at the Fermi surface of FeSe under pressure \cite{PhysRevB.90.144517,PhysRevB.93.094505}. Although it only occupies a small area of the Brillouin zone, it may has a significant impact on the cyclotron frequency of carriers, which is a crucial factor in the emergence and gradual enhancement of the $H$-linear MR.
 SDW fluctuation would be strongly enhanced and significantly affects the transport properties of multi-band material above the magnetic transition \cite{SDW}, including the occurrence of a magnetic tricritical point discussed above \cite{cicuo}. Correspondingly, as reflected in the $\alpha$ cloud diagram in Fig.~\ref{FIG. 1}(b), Hall coefficient \cite{CJG} and NMR evidence \cite{PhysRevB.96.180502}, spin fluctuations may contribute to the appearance of the $H$-linear term. 
 
 In summary, we observed two distinct regions with strange metal behavior in the $P-T$ phase diagram of FeSe. Coincidentally, MR also manifests two $H$-linear areas, and both appear within the strange metal regions. All results and analyzes converge to propose that the $H$-linear MR is closely linked to the strange metal behavior. Furthermore, we confirmed that the two areas of $H$-linear have different origins. 
 The $H$-linear MR under small pressure is caused by high-mobility carriers in the Dirac band within the nematic state. However, according to the scaling, $H$-linear MR at $p \textgreater p_\text{c}$ has different origin from that observed at $p \textless p_\text{c}$.
 Although the $H$-linear MR has different origins or manifestations, we aim to find the possible connection, as they all exist within the strange metal region. This may be the key to a unified understanding of the unusual transport phenomena in unconventional superconductors.

 This work was partly supported by the National Key
 R$\&$D Program of China (Grants No. 2024YFA1408400) and the National Natural Science Fundation of China (Grants No. 12374136, No. 12374135,  No.U1932217, No.12204487).

 X.W. and Y.S. contributed equally to this work.

\bibliography{ref}


\newpage 
\onecolumngrid
\section*{Supplemental Materials} 
\makeatletter
\renewcommand{\thefigure}{S\arabic{figure}}
\makeatother
\setcounter{figure}{0}

\begin{figure}[htbp]
    \centering
    \includegraphics[width=0.4\textwidth]{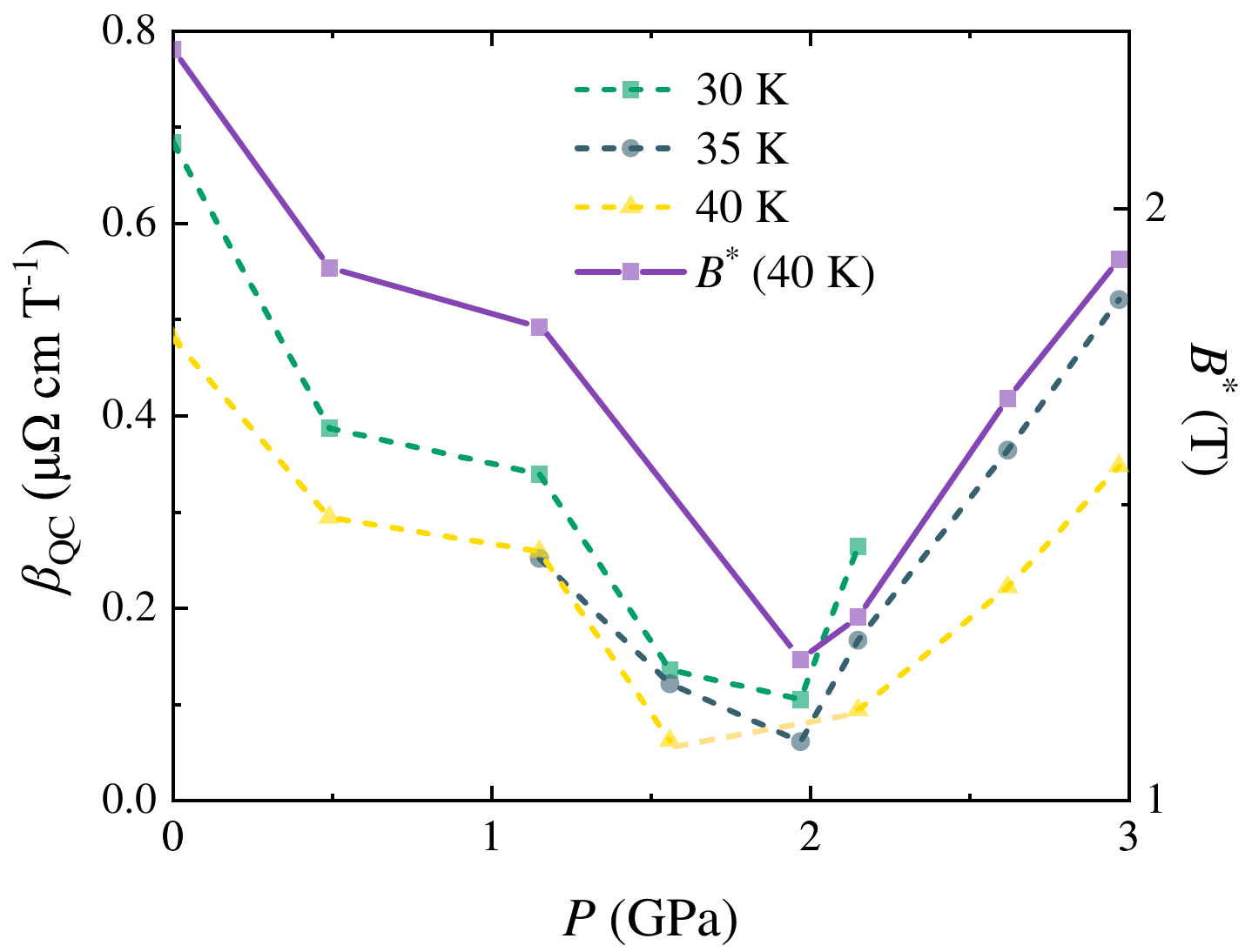}
    \caption{The dashed lines indicate the values of $\beta_\text{QC}$ at different temperatures, while the solid lines represent the intersection points $B$* of the linear and quadratic magnetoresistance.}
\end{figure}

\begin{figure}[htbp]
    \centering
    \includegraphics[width=0.4\textwidth]{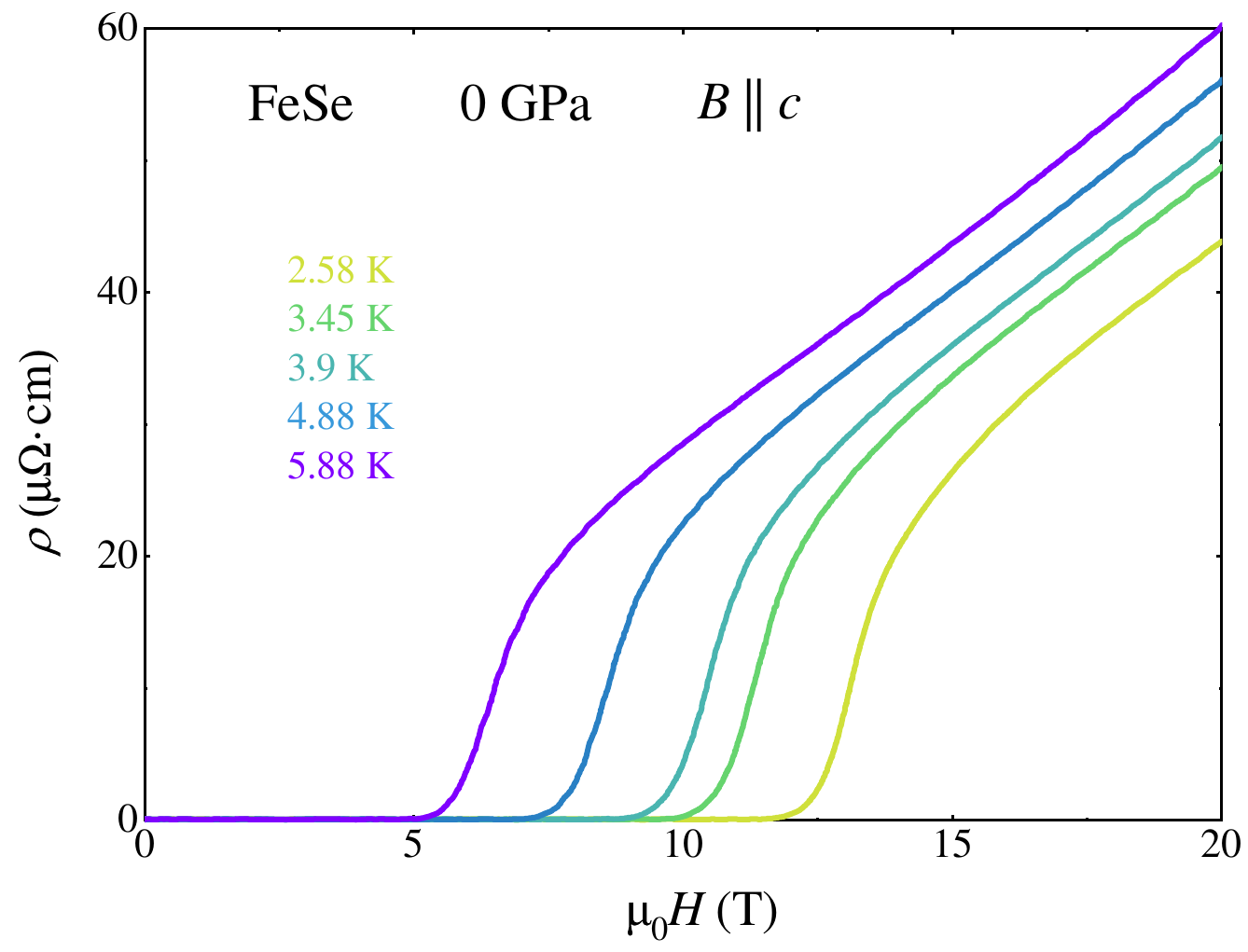}
    \caption{Resistivity versus applied magnetic field from 0 to 20 T for FeSe at different constant temperatures for B||c. In the normal state under high magnetic fields, the $\rho$ - $H$ curves exhibit linear behavior.}
\end{figure}

\begin{figure}[htbp]
    \centering
    \includegraphics[width=0.3\textwidth]{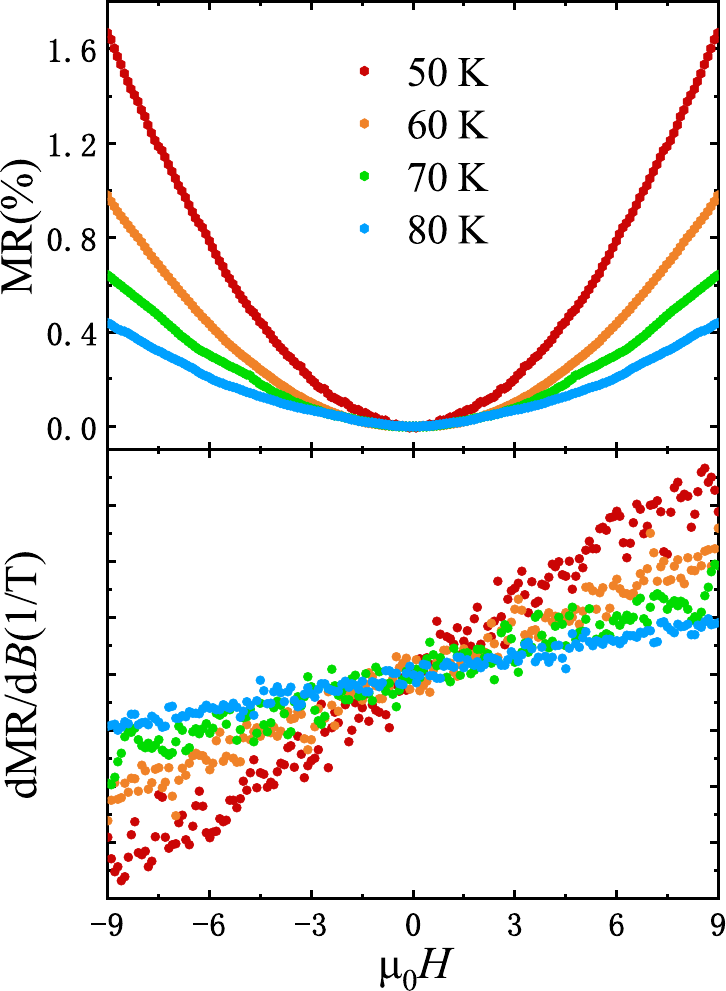}
    \caption{Magnetic field dependence of MR = ($\rho(H)-\rho(0))/\rho(0)$ and field derivative MR [d(MR)/d$B$] for H-FeSe single crystals. In the $H^+$-intercalated FeSe, where $T_\text{S}$ is completely suppressed. Clearly, only containing the quadratic term, MR of the H-FeSe confirms that the H-linear MR is directly related to the nematic order.}
\end{figure}

\end{document}